\documentclass[aps,prc,twocolumn,floatfix,nofootinbib,superscriptaddress]{revtex4-1}

\usepackage[T1]{fontenc}
\usepackage[utf8]{inputenc}
\usepackage{amsmath,amsfonts,amssymb}
\usepackage{graphicx}
\usepackage{bm}
\usepackage{xcolor}
\usepackage{hyperref}
\hypersetup{colorlinks=true,linkcolor=blue,citecolor=blue,urlcolor=blue}

\begin{document}

\title{Nuclear Constraints on $^{12}$C$(\alpha,\gamma)^{16}$O and Their Impact on Black-Hole Mass Predictions}

\author{A.~M.~Mukhamedzhanov}
\affiliation{Texas A\&M University, College Station, Texas 77843, USA}

\begin{abstract}
Gravitational-wave observations have renewed interest in the
black-hole mass gap and in the maximum mass of first-generation black
holes below its lower edge.  The
\(^{12}{\rm C}(\alpha,\gamma)^{16}{\rm O}\) reaction plays a central
role in this problem because it determines the carbon-to-oxygen ratio
after core-helium burning and thereby affects the later evolution of
massive stars toward pulsational pair instability and pair-instability
supernovae.

Recent attempts to constrain \(S(300~{\rm keV})\) from
gravitational-wave population inferences face important limitations,
because the lower edge of the black-hole mass gap is not directly
measured.  It is inferred model dependently from assumptions about
stellar evolution, metallicity, mass loss, rotation, binary evolution,
hierarchical mergers, selection effects, priors, and the adopted
population model.  Therefore, values of \(S(300~{\rm keV})\) inferred
from black-hole populations must remain consistent with independent
nuclear-physics constraints.

In this work we reanalyze the low-energy
\(^{12}{\rm C}(\alpha,\gamma)^{16}{\rm O}\) \(S\) factor using updated
information on the subthreshold \(1^{-}\) and \(2^{+}\) ANCs and on the
ground-state ANC of \(^{16}{\rm O}\), together with direct capture data.
These constraints favor a lower \(S(300~{\rm keV})\) than the older
central evaluation and disfavor very large values required by some
black-hole-population interpretations.  Using the resulting
ANC-constrained \(S(300~{\rm keV})\) range and the transformed relation
between this quantity and the lower edge of the pair-instability mass
gap, we estimate
\[
\frac{M_{\rm BH}}{M_\odot}\simeq 61\text{--}75 .
\]
Thus, the present nuclear-physics constraints favor a relatively high
lower edge of the first-generation black-hole mass gap.
\end{abstract}

\maketitle

\section{Introduction}

The reaction
\(^{12}{\rm C}(\alpha,\gamma)^{16}{\rm O}\) plays a central role in
nuclear astrophysics because it determines the carbon-to-oxygen ratio
established at the end of core-helium burning.  Through its influence on
the CO-core mass and on the subsequent susceptibility of massive stars
to pulsational pair instability (PPI) and pair-instability supernova
(PISN) evolution, this reaction also affects the maximum mass of
first-generation black holes (BHs) below the pair-instability mass gap.
Equivalently, it affects the lower edge of the pair-instability mass
gap produced by stellar collapse.  Recent gravitational-wave population studies have emphasized
the search for the pair-instability black-hole mass gap, whose lower
edge is commonly expected near \(50\)--\(65\,M_\odot\).  This has
motivated attempts to connect the observed BH mass spectrum to the
astrophysical factor \(S(300~{\rm keV})\) for
\(^{12}{\rm C}(\alpha,\gamma)^{16}{\rm O}\)
\cite{Farmer20,Woosley21,Mehta22,Wang25,Tong25,Antonini25,Ray25}.

One of the most important studies in this context is Ref.~\cite{Woosley21}.
Using
\[
S(300~{\rm keV})=140\pm 21~{\rm keV\,b}
\]
from the evaluation of Ref.~\cite{deBoer}, Ref.~\cite{Woosley21}
found that the lower edge of the pair-instability mass gap is shifted to
about \(64\,M_\odot\), while the upper edge is shifted to about
\(161\,M_\odot\).  Rapid rotation can raise the lower edge further, to
approximately \(70\,M_\odot\).  The relevance of this issue has increased because
gravitational-wave observations have reported BHs with masses up to
about \(85\,M_\odot\), as well as numerous BHs in the
\(60\)--\(70\,M_\odot\) range \cite{Abbotta,Abbottb}.  More recently,
Ref.~\cite{Wang25} reported the discovery of new low-spin BHs in the
mass interval \(50\)--\(70\,M_\odot\).

It is important to emphasize that gravitational-wave observations alone
cannot uniquely determine the BH mass gap or the underlying
BH mass spectrum.  The inference depends on the adopted
stellar-evolution model, including nuclear reaction rates, rotation,
mass loss, Eddington-limited accretion, metallicity, and other initial
conditions and parameters.  Among the nuclear reactions that most
strongly affect the BH mass spectrum are the triple-\(\alpha\)
reaction and \(^{12}{\rm C}(\alpha,\gamma)^{16}{\rm O}\).  
The triple-\(\alpha\) rate is often fixed by adopting the
Caughlan--Fowler compilation \cite{Caughlan}.  The astrophysical factor \(S(300~{\rm keV})\)
for \(^{12}{\rm C}(\alpha,\gamma)^{16}{\rm O}\), where
\(300\) keV is the effective energy in the helium-burning Gamow window,
has a particularly important influence on the predicted mass gap.

Regarding the role of
\(^{12}{\rm C}(\alpha,\gamma)^{16}{\rm O}\), we single out
Refs.~\cite{Farmer20,Woosley21,Mehta22}.  These works clearly
demonstrated that the
\(^{12}{\rm C}(\alpha,\gamma)^{16}{\rm O}\) rate strongly influences the
mass of the BH remnant formed in the collapse of a massive star and,
therefore, the boundaries of the pair-instability mass gap.  Depending
on the adopted value of \(S(300~{\rm keV})\), the lower edge of the gap
can vary around $59^{+34}_{-13}\,M_\odot,$
while the upper edge can vary around $139^{+30}_{-14}\,M_\odot.$
These error bars were obtained by assigning a \(\pm3\sigma\) uncertainty
to the \(S(300~{\rm keV})\) factor for
\(^{12}{\rm C}(\alpha,\gamma)^{16}{\rm O}\).  The physical reason is
well understood.  
This reaction controls the post-helium-burning C/O ratio.
A smaller \(S(300~{\rm keV})\) leaves more residual carbon and
therefore increases the C/O ratio,, and helps the star avoid the PPI regime, thereby
allowing a larger first-generation BH remnant.  By contrast, a smaller
post-helium-burning carbon abundance weakens carbon burning and favors
the onset of PPISN or PISN evolution, reducing the maximum BH mass or,
in the case of full PISN, leaving no BH remnant.

Recently, Ref.~\cite{Xin} was accepted for publication.  In that work,
the pair-instability BH mass gap was shifted from
\(104\)--\(184\,M_\odot\) down to approximately
\(45\)--\(135\,M_\odot\).  This implies that the highest BH mass
below the lower edge of the gap would be smaller than
\(\sim45\,M_\odot\).  To support this conclusion, Ref.~\cite{Xin}
appeals to the astrophysical inferences of
Refs.~\cite{Tong25,Antonini25,Wang25,Ray25}.

However, these references do not all support such a low lower-edge mass
cutoff.  Ref.~\cite{Ray25} obtained a lower-edge mass of
\(57^{+17}_{-10}\,M_\odot\), while Ref.~\cite{Wang25} found a cutoff for
low-spin BHs of
\(68^{+19.8}_{-18.5}\,M_\odot\)
at the \(90\%\) credible level, consistent with
\(S(300~{\rm keV})\simeq 110~{\rm keV\,b}\).
Thus, the downward shift advocated in Ref.~\cite{Xin} is supported
mainly by Refs.~\cite{Tong25,Antonini25}, which suggest a lower-edge
mass cutoff near \(40\)--\(50\,M_\odot\).

This interpretation is problematic from the nuclear-physics point of
view.  As discussed in Ref.~\cite{Ray25}, the \(99\%\) credible
intervals associated with Refs.~\cite{Tong25,Antonini25} correspond to
$S(300~{\rm keV})\simeq 150\text{--}450~{\rm keV\,b},$
which only marginally overlaps with the range recommended by
Ref.~\cite{deBoer},
$S(300~{\rm keV})\simeq 130\text{--}160~{\rm keV\,b}.$
Similarly, the mass-gap shift in Ref.~\cite{Xin} to
$\sim45\text{--}135\,M_\odot $
corresponds to
$S(300~{\rm keV})\simeq 138\text{--}263~{\rm keV\,b}.$
The upper part of this interval is not compatible with contemporary
nuclear-physics and nuclear-astrophysics constraints on
\(^{12}{\rm C}(\alpha,\gamma)^{16}{\rm O}\).

Agreement with astrophysical observables alone is therefore not
sufficient.  Any inference based on BH mass distributions must
also remain consistent with independently established nuclear-physics
constraints, including bounds on the ANCs and on the extrapolated
\(S\) factor.  A best fit or posterior maximum that lies outside the
physically allowed nuclear domain is not a physically acceptable
determination; it is only a mathematical compensation for missing,
incomplete, or insufficiently constrained physics.

The purpose of the present work is to emphasize that any program aimed
at determining BH masses and the black-hole mass gap can be
physically meaningful only if the astrophysical inference is aligned
with the independent nuclear constraints that govern the low-energy
\(^{12}{\rm C}(\alpha,\gamma)^{16}{\rm O}\) capture amplitude.  Many
studies pursuing this program have relied heavily on the state-of-the-art
review by deBoer {\it et al.} \cite{deBoer}, published in 2017.  Since
then, however, new results on the involved ANCs have become available that tend to shift the
range
\[
S(300~{\rm keV})=130\text{--}160~{\rm keV\,b}
\]
from Ref.~\cite{deBoer} toward lower values.  A decrease of
\(S(300~{\rm keV})\) lowers the
\(^{12}{\rm C}(\alpha,\gamma)^{16}{\rm O}\) reaction rate, increases the
maximum black-hole mass below the gap, and shifts the lower edge of the
mass gap upward.  These conclusions are based on new information about
the asymptotic normalization coefficients (ANCs) that control the
low-energy \(S\) factor.  In what follows, we discuss this new ANC
information and its impact on black-hole masses below the lower edge of
the pair-instability mass gap.

All calculations reported below were performed within the multilevel
\(R\)-matrix formalism using the Brune parametrization, following the
approach adopted in Ref.~\cite{deBoer}.

\section{Subthreshold  and  ground-state ANCs} 

At low energies, the $^{12}\mathrm{C}(\alpha,\gamma)^{16}\mathrm{O}$ radiative-capture amplitude is dominated by the near-threshold subthreshold states of ${}^{16}\mathrm{O}$: the $1^{-}$ state at $7.12$ MeV and the $2^{+}$ state at $6.92$ MeV. 
These states contribute through their subthreshold resonances, while
additional contributions arise from resonances above threshold,
direct capture (for \(E2\)), and smaller cascade transitions through
bound excited states \cite{muk24}.  The subthreshold resonance contributions to the radiative capture arise through the long-range tails of their bound-state wave functions in the external region. 
The strengths of the subthreshold-resonance contributions are determined
by the corresponding ANCs, $C_{1}$  and $C_{2}$, determining the amplitudes of the overlap functions for the virtual decays ${}^{16}{\rm O}(7.12\,{\rm MeV},\,1^{-})$  and  ${}^{16}{\rm O}(6.92\,{\rm MeV},\,2^{+}),$   respectively.
Together with the ground-state ANC
\(C_{0}\equiv{}^{16}{\rm O}(0^{+};\,0.0\,{\rm MeV})\),
the subthreshold ANCs control the
\(S(300\,{\rm keV})\) factor.  Their quantitative contributions depend on the adopted ANC values. Table I of \cite{muk26}  presents a snapshot of the available  experimental and theoretical ANC values.

There have been several important developments concerning the
subthreshold and ground-state ANCs since the publication of
Ref.~\cite{deBoer}.  First, these developments concern the subthreshold
ANCs \(C_{1}\) and \(C_{2}\).  Low values of these ANCs were extracted
from the sub-Coulomb transfer reaction
\begin{align}
{}^{12}\mathrm{C}({}^{6}\mathrm{Li},d){}^{16}\mathrm{O}
\label{subCoultrreact1}
\end{align}
\cite{Brune,Avila}. These ANC determinations are regarded as among the most reliable
currently available because the sub-Coulomb transfer reaction is
practically insensitive to nuclear distortions in the initial and final
channels. In addition, the \(Q\) value of the reaction is very small and the
reaction is extremely peripheral.  Therefore, no additional adjustable
parameters, apart from the involved ANCs, are required to calculate the
sub-Coulomb transfer differential cross section.  The differential cross
section is proportional to
\[
\frac{d\sigma}{d\Omega}\propto
C_{\alpha\,{}^{12}\mathrm{C}}^{2}\,C_{\alpha d}^{2}.
\]

In Refs.~\cite{Brune,Avila}, the extraction of \(C_{1}\) and \(C_{2}\)
used the \({}^{6}{\rm Li}\) ANC
\[
C_{\alpha d}=5.3\pm0.5~\mathrm{fm}^{-1/2}
\]
from Ref.~\cite{Blokhintsev93}.  

The subthreshold ANCs extracted from the sub-Coulomb transfer reaction
(\ref{subCoultrreact1}),
\[
C_{1}=2.08\times10^{14}~{\rm fm}^{-1/2},
\qquad
C_{2}=1.14\times10^{5}~{\rm fm}^{-1/2},
\]
were regarded as among the most reliable and were adopted in
Ref.~\cite{deBoer} for the \(R\)-matrix analysis of the
\(^{12}{\rm C}(\alpha,\gamma)^{16}{\rm O}\) reaction.

More recently,
\emph{ab initio} no-core shell-model-with-continuum calculations
revised the value of \(C_{\alpha d}^{2}\) upward by about \(30\%\)\cite{Hebborn},
\[
C_{\alpha d}^{2}=6.864\pm0.210~\mathrm{fm}^{-1},
\]
which implies a reduction of about \(14\%\) in the deduced values of
\(C_{1}\) and \(C_{2}\).  The revised ANCs become
\begin{align}
&C_{1}= (1.83 \pm 0.08)\times 10^{14}~\mathrm{fm}^{-1/2}, \nonumber\\
&C_{2}= (0.98 \pm 0.08)\times 10^{5}~\mathrm{fm}^{-1/2}.
\label{C1C2Hebborn}
\end{align}
These revised values are significantly lower than those adopted in
Ref.~\cite{deBoer} and therefore tend to reduce the extrapolated
low-energy \(E1\) and \(E2\) contributions.
Since the capture through a subthreshold state is proportional to
\(C_{s}^{2}\) (\(C_{s}=C_{1},\,C_{2}\)), the reduction of the
subthreshold \(C_{s}^{2}\) values implies a comparable reduction in the
corresponding subthreshold contribution to
\(S(300\,{\rm keV})\).

Note that recently determined subthreshold ANCs obtained by
extrapolating elastic-scattering phase shifts to the corresponding
bound-state poles \cite{Blokhintsev23},
$C_{1}=(1.42 \pm 0.05)\times 10^{14}~\mathrm{fm}^{-1/2}$
and 
$C_{2}=(2.27 \pm 0.02)\times 10^{5}~\mathrm{fm}^{-1/2}$
lie near the upper end of the adopted range of subthreshold ANCs.
 However,
their accuracy is limited by the uncertainties in the elastic-scattering
phase shifts, although the extrapolation method itself is model
independent.

The most significant change in the ANC landscape since the evaluation of
Ref.~\cite{deBoer} concerns the ground-state ANC \(C_0\).  As summarized
in Table~\ref{Table_C0}, all recent studies included in the present
comparison give values
\[
C_0 \gtrsim 600~{\rm fm}^{-1/2},
\]
in contrast to the value
\[
C_0=58~{\rm fm}^{-1/2}
\]
adopted in Ref.~\cite{deBoer}.  This change has important consequences
for the external-capture contribution to the ground-state transition and
therefore for the extrapolated \(S\) factor at astrophysical energies.

\begin{table*}[t] 
\caption{
ANC values \(C_0\) for the overlap
\(^{16}{\rm O}(\mathrm{g.s.})\to \alpha+{}^{12}{\rm C}(\mathrm{g.s.})\).
For Ref.~\cite{Morais}, three values of \(C_0\) were obtained using
three different interaction potentials.}
\label{Table_C0}
\centering
\renewcommand{\arraystretch}{1.15}
\setlength{\tabcolsep}{5pt}
\begin{tabular}{|c|c|c|c|c|c|c|}
\hline
ANC &  Ref.~\cite{Kundalia} &  Ref.~\cite{Sayre}  & Ref.~\cite{Chien}
& Ref.~\cite{Adhikari} & Ref.~\cite{Morais} & Ref.~\cite{Blokhintsev26} \\
\hline
\(C_0\) (fm\(^{-1/2}\)) &
\(709\) &
\(744\pm144\) &
\(740\) &
\(637\pm86\) &
\(3390;\ 1230;\ 790\) &
\(837\text{--}841\) \\
\hline
\end{tabular}
\end{table*}

\section{Bayesian constraint on \(C_0\) from the \(E2\) component}
\label{sec:SE2BayesC0}

The direct \(E1\) transition to the ground state of
\(^{16}{\rm O}\) in the reaction
\(^{12}{\rm C}(\alpha,\gamma)^{16}{\rm O}\) is strongly suppressed
because it is isospin forbidden.  Consequently, the direct \(E1\)
external-capture contribution to the ground-state transition is
negligible.  The \(E1\) contribution is therefore governed mainly by the
resonant and subthreshold \(1^-\) amplitudes and does not provide a
useful direct constraint on the ground-state ANC \(C_0\).  For this
reason, in selecting the most probable value of \(C_0\), we begin with
the \(E2\) component,  where the external-capture contribution to the
ground-state transition is more directly connected with \(C_0\).

Guided by the available experimental information summarized in
Table~\ref{Table_C0}, we perform a Bayesian analysis using a flat prior
in the interval
\[
600\le C_0\le 900~{\rm fm}^{-1/2}.
\]
The extracted \(S_{E2}\) data of Ref.~\cite{Assunc} are used to constrain
\(C_0\), while the subthreshold \(2^+\) ANC is kept fixed at the value
favored by the sub-Coulomb transfer analysis, see Eq. (\ref{C1C2Hebborn}).
This is an important constraint because \(C_2\) controls the strength of
the subthreshold \(2^+\) contribution and the associated interference
pattern in the \(E2\) capture amplitude.  

In contrast, the dependence 
on the ground-state ANC \(C_0\) is weaker.
This is because the dominant \(E2\) contribution at low energy arises
mainly from the interference between the subthreshold \(2^{+}\) resonance
and the direct-capture amplitude to the ground state.  The ground-state
ANC \(C_0\) enters primarily through the normalization of this external
direct-capture amplitude and therefore has a less pronounced effect than
the subthreshold ANC \(C_2\).

For each selected value of \(C_0\), the \(E2\) capture calculation was
repeated and compared with the extracted \(S_{E2}\) data.  The likelihood
was taken in the Gaussian form,
\[
{\cal L}(D_{E2}|C_0)
\propto
\exp\left[-\frac{1}{2}\chi^2_{E2}(C_0)\right],
\]
where
\[
\chi^2_{E2}(C_0)
=
\sum_i
\left[
\frac{
S_{E2}^{\rm th}(E_i;C_0,C_2)-S_{E2,i}^{\rm exp}
}{
\sigma_i
}
\right]^2 .
\]
Here \(D_{E2}\) denotes the set of extracted \(E2\) data points used in
the fit.  The quantity \({\cal L}(D_{E2}|C_0)\) is the likelihood function,
which measures how well a given value of \(C_0\) reproduces the data
\(D_{E2}\).  The theoretical value
\(S_{E2}^{\rm th}(E_i;C_0,C_2)\) is the calculated \(E2\) astrophysical
\(S\) factor at the experimental energy \(E_i\), obtained with the trial
value of \(C_0\) and the fixed subthreshold ANC \(C_2\).  The quantity
\(S_{E2,i}^{\rm exp}\) is the corresponding experimentally extracted
\(E2\) astrophysical \(S\) factor, and \(\sigma_i\) is its quoted
experimental uncertainty.

Thus, each term in the sum is the squared residual between theory and
experiment, measured in units of the experimental error.  The exponential
form of the likelihood follows from assuming Gaussian, uncorrelated
experimental uncertainties.  Under this assumption, smaller values of
\(\chi^2_{E2}(C_0)\) correspond to larger likelihoods and therefore to
larger posterior probabilities.

The flat prior is
\[
\pi(C_0)=
\begin{cases}
{\rm const}, & 600\le C_0\le 900~{\rm fm}^{-1/2},\\
0, & \text{otherwise}.
\end{cases}
\]
The posterior distribution is then
\[
P(C_0|D_{E2})
=
\frac{
{\cal L}(D_{E2}|C_0)\,\pi(C_0)
}{
\int {\cal L}(D_{E2}|C_0)\,\pi(C_0)\,dC_0
}.
\]
Equivalently, for numerical stability, we evaluate
\[
P(C_0|D_{E2})
\propto
\exp\left[
-\frac{1}{2}
\left(
\chi^2_{E2}(C_0)-\chi^2_{E2,\min}
\right)
\right],
\]
inside the prior interval.  Here \(\chi^2_{E2,\min}\) denotes the
minimum value of \(\chi^2_{E2}(C_0)\) obtained within the scanned prior
interval.  Thus
\[
\chi^2_{E2}(C_0)-\chi^2_{E2,\min}
\equiv
\Delta\chi^2_{E2}(C_0),
\]
and the subtraction only rescales the likelihood by a constant factor
without changing the posterior shape.

\begin{figure}[t]
\centering
\includegraphics[width=0.9\linewidth]{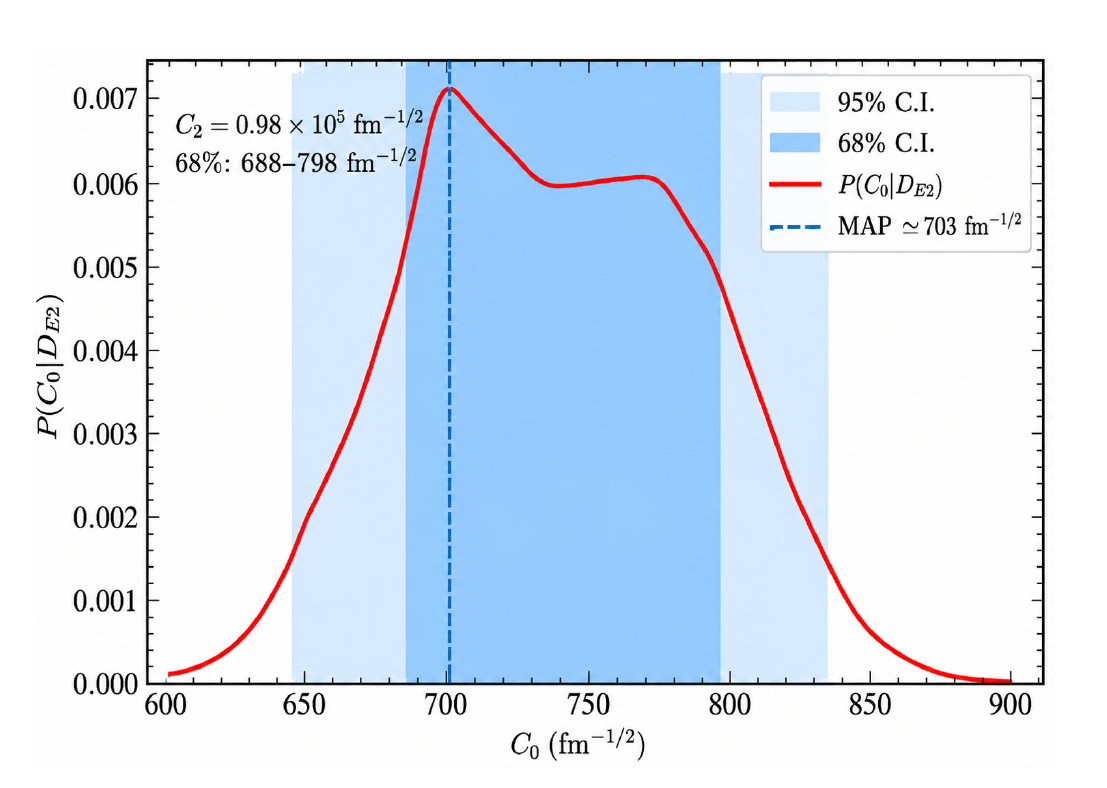}
\caption{
Smoothed Bayesian posterior distribution for the ground-state ANC
\(C_0\), obtained from the extracted \(S_{E2}\) data with the
subthreshold \(2^+\) ANC fixed at
\(C_2=0.98\times10^5~{\rm fm}^{-1/2}\).  A flat prior was used over the
interval \(600\le C_0\le900~{\rm fm}^{-1/2}\).  The posterior peaks at
\(C_0^{\rm MAP}\simeq703~{\rm fm}^{-1/2}\), while the posterior median is
\(C_0^{\rm med}\simeq740~{\rm fm}^{-1/2}\).
 The shaded regions denote
the 68\% and 95\% credible intervals,
\(688\le C_0\le798~{\rm fm}^{-1/2}\) and
\(645\le C_0\le840~{\rm fm}^{-1/2}\), respectively.
}
\label{fig:PosteriorC0paper}
\end{figure}

The resulting posterior distribution is shown in
Fig.~\ref{fig:PosteriorC0paper}.\footnote{
The vertical axis shows the posterior probability density, not the
probability itself.  The density is normalized according to
\(\int_{600}^{900} P(C_0|D_{E2})\,dC_0=1\).  Since \(C_0\) is measured in
fm\(^{-1/2}\) and the posterior extends over a range of order
\(10^2~{\rm fm}^{-1/2}\), density values of order
\(10^{-3}\)--\(10^{-2}\) are expected.
}

The maximum of the posterior, or MAP value, occurs at
$C_0^{\rm MAP}=703~{\rm fm}^{-1/2}.$
The median value is $C_0^{\rm med}\simeq 740~{\rm fm}^{-1/2},$
and the credible intervals are
\[
688\le C_0\le 798~{\rm fm}^{-1/2}
\qquad (68\%),
\]
and
\[
645\le C_0\le 840~{\rm fm}^{-1/2}
\qquad (95\%).
\]
Thus the Bayesian analysis favors the same range of \(C_0\) values
indicated by recent ANC determinations summarized in
Table~\ref{Table_C0}.\footnote{
We do not include the result of Ref.~\cite{Shen}, which reported
\(C_0=337\pm45~{\rm fm}^{-1/2}\), because it lies outside the
95\% credible interval obtained in the present \(S_{E2}\)-only Bayesian
analysis and is also in tension with the other recent determinations
listed in Table~\ref{Table_C0}.
}

The corresponding \(E2\) contribution at \(E=300\) keV is approximately
$S_{E2}(300\,{\rm keV})=50~{\rm keV\,b}$
for $C_0^{\rm med}=740~{\rm fm}^{-1/2},$
and $S_{E2}(300\,{\rm keV})=49~{\rm keV\,b}$
for the posterior maximum,
$C_0^{\rm MAP}=703~{\rm fm}^{-1/2}.$
Over the main posterior region the variation of \(S_{E2}(300\,{\rm keV})\) with
\(C_0\) is very modest.  For example, increasing \(C_0\) from \(740\) to
\(870~{\rm fm}^{-1/2}\) changes \(S_{E2}(300~{\rm keV})\) only by a few
 keV\,b.
This confirms that the low-energy \(E2\) extrapolation is more sensitive
to the subthreshold ANC \(C_2\) and to the interference pattern than to
moderate variations of \(C_0\).

The posterior also provides a useful lower-side constraint.  When the
scan was extended below the prior interval, the case
\(C_0=550~{\rm fm}^{-1/2}\) gave a substantially worse description of
the extracted \(S_{E2}\) data.  The best interference branch for
\(C_0=550~{\rm fm}^{-1/2}\) yielded
\[
\chi^2_{E2}=22.13,
\qquad 
\chi^2_{\nu,E2}=2.77,
\]
compared with
\[
\chi^2_{E2}\simeq 5.8,
\qquad
\chi^2_{\nu,E2}\simeq 0.73,
\]
near the preferred region around
\(C_0^{\rm MAP}=703~{\rm fm}^{-1/2}\).  For the median value
\(C_0^{\rm med}=740~{\rm fm}^{-1/2}\), we find
\[
\chi^2_{\nu,E2}\simeq 0.78.
\]
Therefore, the \(S_{E2}\)-only data do not favor a drift toward very low
values of the ground-state ANC.

We emphasize that the \(S_{E2}\)-only posterior is used as the cleaner
Bayesian diagnostic for the \(E2\) extrapolation, while the total-capture
comparison serves as a consistency test  of the complete
\(E1+E2+\)cascade model.

\section{ Comparison with experimental data}
\label{sec:comparison}

We present here the comparison of the three-level \(R\)-matrix
calculations for the \(E1\) and \(E2\) radiative transitions in the
\(^{12}{\rm C}(\alpha,\gamma)^{16}{\rm O}\) reaction.  As discussed
above, for this comparison we use the extracted \(E1\) and \(E2\)
multipole data of Ref.~\cite{Assunc}.  We adopt
\[
C_1=1.83\times10^{14}~{\rm fm}^{-1/2},\qquad
C_2=0.98\times10^{5}~{\rm fm}^{-1/2},
\]
which are the most reliable latest subthreshold ANCs, and the posterior median
value
\[
C_0^{\rm med}=740~{\rm fm}^{-1/2}
\]
obtained from the Bayesian analysis.  These values are used in the
following calculations.

\begin{figure}[t]
\centering
\includegraphics[width=0.6\linewidth,angle=-90]{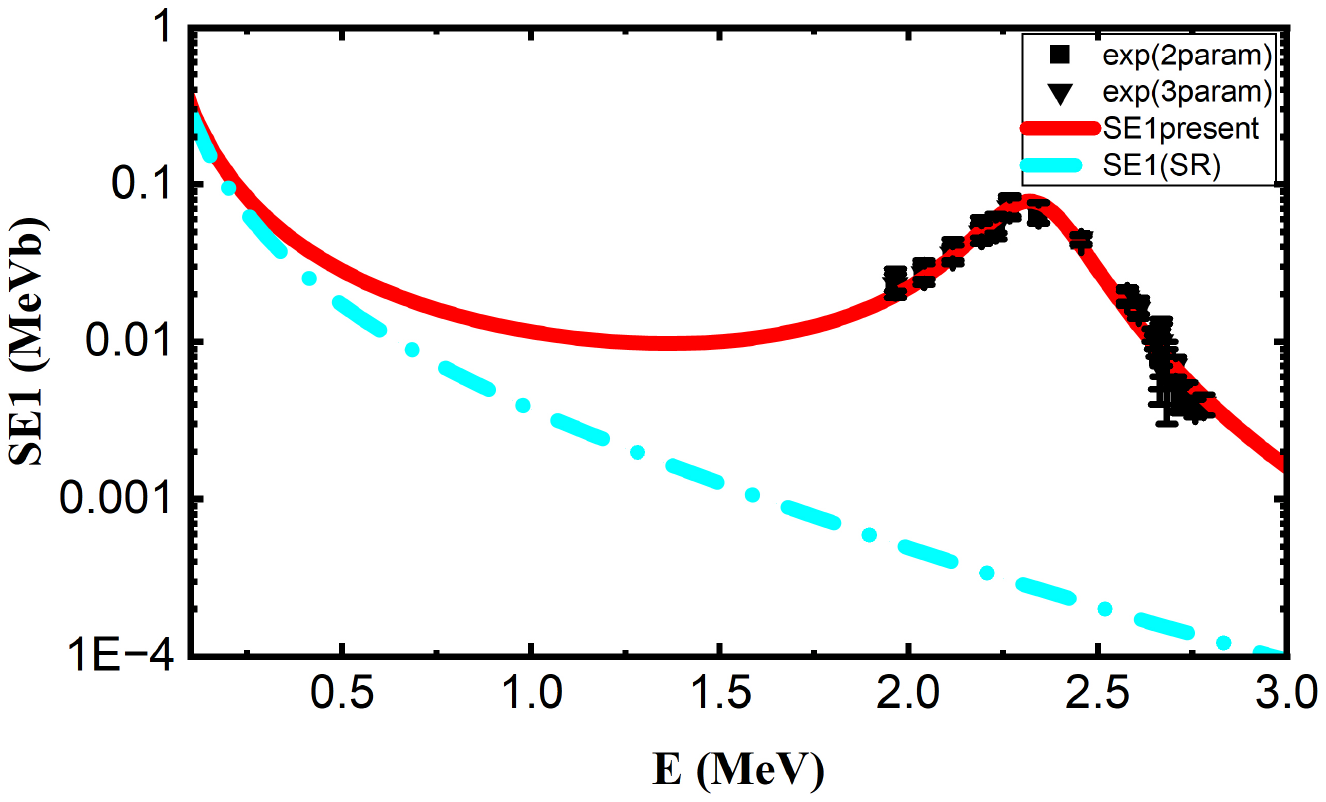}
\caption{
The \(S_{E1}(E)\) factor obtained from the \(R\)-matrix analysis,
compared with the experimental two-parameter and three-parameter data of
Ref.~\cite{Assunc}.  The black rectangles are the two-parameter data,
and the inverted black triangles are the three-parameter data.  The red
solid line is the best \(R\)-matrix fit.  The green dash-dotted line is
the contribution from the \(1^{-}\) subthreshold resonance.
}
\label{fig_SE1}
\end{figure}

\begin{figure}[t]
\centering
\includegraphics[width=0.7\linewidth,angle=-90]{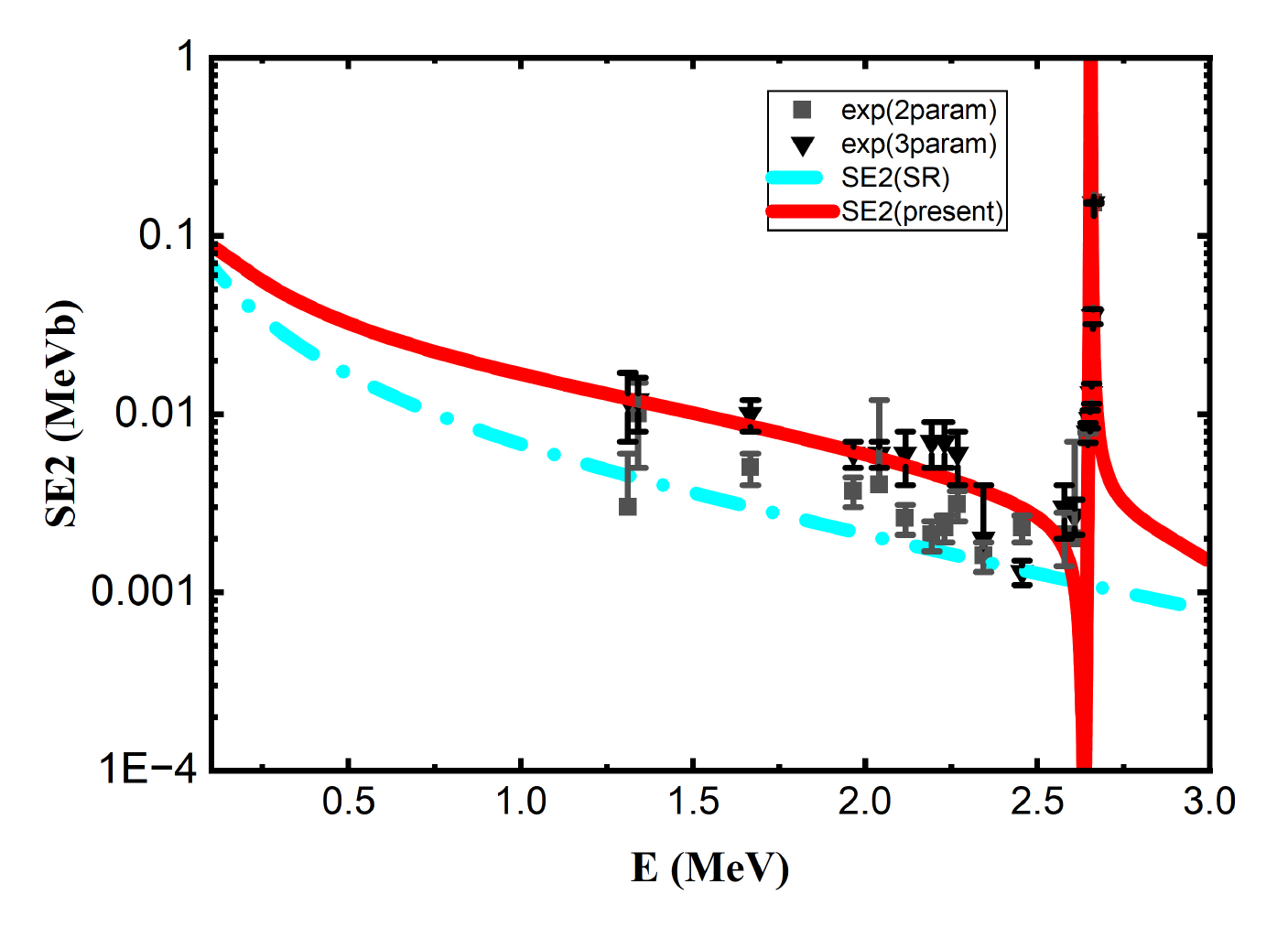}
\caption{
The \(S_{E2}(E)\) factor obtained from the \(R\)-matrix analysis,
compared with the experimental two-parameter and three-parameter data of
Ref.~\cite{Assunc}.  The notation is the same as in
Fig.~\ref{fig_SE1}, except that the green dash-dotted line shows the
contribution from the \(2^{+}\) subthreshold resonance.
}
\label{fig_SE2}
\end{figure}

In Figs.~\ref{fig_SE1} and \ref{fig_SE2}, the calculated and
experimental \(S\) factors are shown. The adopted fitting
parameters are listed in Tables~\ref{tab:SE1} and \ref{tab:SE2}.
The calculated \(S\) factors
were fitted to the experimental multipole data using the multilevel
\(R\)-matrix approach following Ref.~\cite{deBoer}. 

In both figures the red solid lines are the calculated total
\(S_{E1}\) and \(S_{E2}\) factors, while the green dash-dotted curves
show the contributions from the subthreshold resonances, \(1^{-}\) for
\(E1\) and \(2^{+}\) for \(E2\).  These contributions are governed by the corresponding subthreshold ANCs and give significant fractions of the total \(S_{E1}(300~{\rm keV})\) and
\(S_{E2}(300~{\rm keV})\) factors, as summarized in
Tables~\ref{Table_SetsANCs} and  \ref{Table_SfactorsSets}.  Figures  ~\ref{fig_SE1} and \ref{fig_SE2}  illustrate why
the subthreshold ANCs \(C_1\) and \(C_2\) play a pivotal role in
determining \(S_{E1}(300~{\rm keV})\) and
\(S_{E2}(300~{\rm keV})\), respectively.   

\begin{table}[t]
\centering
\small
\caption{Definitions of the parameters adopted in the \(R\)-matrix fit
for the \(E1\) transition.}
\label{tab:SE1}
\begin{tabular}{|l|p{0.58\columnwidth}|}
\hline
Parameter & Description \\
\hline
\(r_0=5.43~{\rm fm}\) 
& Channel radius used in the \(R\)-matrix calculation. \\
\hline
\(\epsilon_{s1}=0.045~{\rm MeV}\) 
& Binding energy of the \(1^{-}\) subthreshold state relative to the
\(\alpha+{}^{12}{\rm C}\) threshold. \\
\hline
\(E_{s1}=-0.045~{\rm MeV}\) 
& Energy of the \(1^{-}\) subthreshold state measured relative to the
\(\alpha+{}^{12}{\rm C}\) threshold. \\
\hline
\(\epsilon_0=7.162~{\rm MeV}\) 
& Binding energy of the \(^{16}{\rm O}\) ground state relative to the
\(\alpha+{}^{12}{\rm C}\) threshold. \\
\hline
\(E_{R1}=2.423~{\rm MeV}\) 
& Resonance energy of the first above-threshold \(1^{-}\) state. \\
\hline
\(E_{RBG}=10.5~{\rm MeV}\) 
& Energy of the background \(1^{-}\) resonance. \\
\hline
\(C_0=740~{\rm fm}^{-1/2}\) 
& ANC of the \(^{16}{\rm O}\) ground state. \\
\hline
\(C_1=1.83\times10^{14}~{\rm fm}^{-1/2}\) 
& ANC of the \(1^{-}\) subthreshold state. \\
\hline
\(\Gamma_{R1}=0.50~{\rm MeV}\) 
& Particle width of the first above-threshold \(1^{-}\) resonance. \\
\hline
\(\Gamma_{RBG}=1.0~{\rm MeV}\) 
& Particle width of the background \(1^{-}\) resonance. \\
\hline
\(\Gamma_{\gamma 0}=55\times10^{-9}~{\rm MeV}\) 
& Radiative width of the \(1^{-}\) subthreshold state for the transition
to the ground state. \\
\hline
\(\Gamma_{\gamma R1}=19\times10^{-9}~{\rm MeV}\) 
& Radiative width of the first above-threshold \(1^{-}\) resonance for
the transition to the ground state. \\
\hline
\(\Gamma_{\gamma BG0}=1.4\times10^{-6}~{\rm MeV}\) 
& Radiative width of the background \(1^{-}\) resonance for the
transition to the ground state. \\
\hline
\(L=1\) 
& Electromagnetic multipolarity of the \(E1\) transition. \\
\hline
\(\ell_i=1\) 
& Initial-channel orbital angular momentum. \\
\hline
\(\ell_f=0\) 
& Final bound-state orbital angular momentum for the ground-state
transition. \\
\hline
\(J_i=1\) 
& Spin of the initial \(1^{-}\) state. \\
\hline
\(J_f=0\) 
& Spin of the final ground state of \(^{16}{\rm O}\). \\
\hline
\end{tabular}
\end{table}

The fit for the \(E1\) transition is more straightforward than for
\(E2\), because the direct \(E1\) transition to the ground state is
negligible and \(S_{E1}\) is practically insensitive to the value of
\(C_0\).  The dominant contribution to
\(S_{E1}(300~{\rm keV})\) comes from the subthreshold \(1^{-}\) state and
the above-threshold \(1^{-}\) resonance at \(E_{R1}=2.423\) MeV.  To fit
the experimental \(E1\) data of Ref.~\cite{Assunc}, we adopted a
three-level \(R\)-matrix model including the subthreshold \(1^{-}\)
state at \(E_{s1}=-0.045\) MeV, the resonance at
\(E_{R1}=2.423\) MeV, and a background resonance.  As in
Ref.~\cite{deBoer}, the subthreshold state and the $2.423$  MeV resonance
interfere constructively.  The background resonance contributes mainly
at higher energies and does not affect the low-energy behavior near
$300$ keV.  The best fit gives
\[
S_{E1}(300~{\rm keV})=62~{\rm keV\,b}
\]
with reduced chi-squares \(\chi_\nu^2=1.35\) for the two-parameter data
set and \(\chi_\nu^2=1.15\) for the three-parameter data set.

The fit for the \(E2\) transition is more delicate.
To fit the \(E2\) data of Ref.~\cite{Assunc}, we used, as in
Ref.~\cite{deBoer}, a three-level \(R\)-matrix model containing the
subthreshold \(2^{+}\) state, the narrow \(2^{+}\) resonance at
\(E_{R1}=2.654\) MeV, and the broader \(2^{+}\) resonance at
\(E_{R2}=4.344\) MeV.  The dominant low-energy contribution to
\(S_{E2}(300~{\rm keV})\) comes from the subthreshold \(2^{+}\) state
and from direct capture to the \(^{16}{\rm O}\) ground state.

\begin{table}[t]
\centering
\small
\setlength{\tabcolsep}{4pt}
\caption{Definitions of the parameters adopted in the \(R\)-matrix fit
for the \(E2\) transition.}
\label{tab:SE2}
\begin{tabular}{|l|p{0.58\columnwidth}|}
\hline
Parameter & Description \\
\hline
\(r_0=5.43~{\rm fm}\) 
& Channel radius used in the \(R\)-matrix calculation. \\
\hline
\(\epsilon_{s2}=0.2449~{\rm MeV}\) 
& Binding energy of the \(2^{+}\) subthreshold state relative to the
\(\alpha+{}^{12}{\rm C}\) threshold. \\
\hline
\(E_{s2}=-0.2449~{\rm MeV}\) 
& Energy of the \(2^{+}\) subthreshold state measured relative to the
\(\alpha+{}^{12}{\rm C}\) threshold. \\
\hline
\(\epsilon_0=7.162~{\rm MeV}\) 
& Binding energy of the \(^{16}{\rm O}\) ground state relative to the
\(\alpha+{}^{12}{\rm C}\) threshold. \\
\hline
\(E_{R1}=2.654~{\rm MeV}\) 
& Resonance energy of the first above-threshold \(2^{+}\) state. \\
\hline
\(E_{R2}=4.344~{\rm MeV}\) 
& Resonance energy of the second above-threshold \(2^{+}\) state. \\
\hline
\(C_0=740~{\rm fm}^{-1/2}\) 
& ANC of the \(^{16}{\rm O}\) ground state. \\
\hline
\(C_2=0.98\times10^5~{\rm fm}^{-1/2}\) 
& ANC of the \(2^{+}\) subthreshold state. \\
\hline
\(\Gamma_{R1}=6.2\times10^{-4}~{\rm MeV}\) 
& Particle width of the first above-threshold \(2^{+}\) resonance. \\
\hline
\(\Gamma_{R2}=0.12~{\rm MeV}\) 
& Particle width of the second above-threshold \(2^{+}\) resonance. \\
\hline
\(\Gamma_{\gamma 0} = 9.7\times 10^{-8}~{\rm MeV}\)
& Radiative width of the \(2^{+}\) subthreshold state for the transition
to the ground state. \\
\hline
\(\Gamma_{\gamma R1}=5\times10^{-9}~{\rm MeV}\)
& Radiative width of the first above-threshold \(2^{+}\) resonance for
the transition to the ground state. \\
\hline
\(\Gamma_{\gamma R2}=6.2\times10^{-7}~{\rm MeV}\) 
& Radiative width of the second above-threshold \(2^{+}\) resonance for
the transition to the ground state. \\
\hline
\(L=2\) 
& Electromagnetic multipolarity of the \(E2\) transition. \\
\hline
\(\ell_i=2\) 
& Initial-channel orbital angular momentum for the \(E2\) transition. \\
\hline
\(\ell_f=0\) 
& Final bound-state orbital angular momentum for the ground-state
transition. \\
\hline
\(J_i=2\) 
& Spin of the initial \(2^{+}\) state. \\
\hline
\(J_f=0\) 
& Spin of the final ground state of \(^{16}{\rm O}\). \\
\hline
\end{tabular}
\end{table}

The large value \(C_0=740~{\rm fm}^{-1/2}\) significantly complicates
the \(E2\) fit.  In Ref.~\cite{deBoer}, the combination
\(C_2=1.14\times10^5~{\rm fm}^{-1/2}\) and the much smaller
\(C_0=58~{\rm fm}^{-1/2}\) favored destructive interference between the
subthreshold \(2^+\) amplitude and the direct-capture amplitude to the
ground state.  In contrast, when the larger modern value of \(C_0\) is
adopted together with the lower subthreshold ANC \(C_2\), the fit
requires constructive interference between the subthreshold and direct
capture amplitudes.  With this choice, we obtain a good description of
the three-parameter \(E2\) data of Ref.~\cite{Assunc}; the two-parameter
data lie systematically lower and are not fitted simultaneously.  The
best fit has $\chi_\nu^2=0.78,$  and gives
$S_{E2}(300~{\rm keV})=50~{\rm keV\,b}.\,$
In evaluating \(\chi_\nu^2\), we excluded the narrow-resonance region
around \(E_{R1}\) and the point at \(E=2.455\) MeV, which lies visibly
below the surrounding data and has a very small uncertainty, thereby
driving the total \(\chi_\nu^2\) to an unrealistically large value.  The
parameter set of Ref.~\cite{deBoer} also gives a good fit, with
\(\chi_\nu^2=0.96\) and
\[
S_{E2}(300~{\rm keV})=51~{\rm keV\,b}.
\]

\begin{table}[htb]
\caption{ANC sets used in the calculations summarized in
Table~\ref{Table_SfactorsSets}.}
\label{Table_SetsANCs}
\centering
\begin{tabular}{|c|c|c|c|}
\hline
Set & \(C_1\) & \(C_2\) & \(C_0\) \\
& (\({\rm fm}^{-1/2}\)) & (\({\rm fm}^{-1/2}\)) & (\({\rm fm}^{-1/2}\)) \\
\hline
1 & \(1.83\times10^{14}\) & \(0.98\times10^{5}\) & \(740\) \\
2 & \(2.08\times10^{14}\) & \(1.14\times10^{5}\) & \(740\) \\
\hline
\end{tabular}
\end{table}

\begin{table}[htb]
\caption{
The \(S(300~{\rm keV})\) factors for the radiative capture reaction
\(^{12}{\rm C}(\alpha,\gamma)^{16}{\rm O}\).  Here \(S_{E1}\) and
\(S_{E2}\) are the total \(S\) factors for the \(E1\) and \(E2\)
transitions, respectively, and \(S_{\rm tot}=S_{E1}+S_{E2}\).  The
quantities \(S_{E1}^{({\rm SR})}\) and \(S_{E2}^{({\rm SR})}\) denote
the subthreshold-resonance contributions at \(300\) keV.  All \(S\)
factors are in keV b.  Cascade transitions are not included.
}
\label{Table_SfactorsSets}
\centering
\begin{tabular}{|c|c|c|c|c|c|c|c|}
\hline
Set &
\(S_{E1}\) &
\(S_{E2}\) &
\(S_{\rm tot}\) &
\(S_{E1}^{({\rm SR})}\) &
\(S_{E2}^{({\rm SR})}\) &
\(\dfrac{S_{E1}^{({\rm SR})}}{S_{E1}}\) &
\(\dfrac{S_{E2}^{({\rm SR})}}{S_{E2}}\) \\
\hline
1 & 62 & 50 & 112 & 46 & 30 & 0.74 & 0.60 \\
2 & 85 & 58 & 143 & 60 & 46 & 0.71 & 0.79 \\
\hline
\end{tabular}
\end{table}

Table~\ref{Table_SetsANCs} displays two ANC sets used in the
calculations.  Set 1 is the present choice, based on the updated subthreshold ANCs and
the Bayesian median value adopted for the modern large ground-state ANC
\(C_0\).  Set 2 keeps the original subthreshold ANCs of Refs.~\cite{Brune,Avila}, as used in
Ref.~\cite{deBoer}, but adopts the same large value
\(C_0=740~{\rm fm}^{-1/2}\).  This comparison isolates the effect of the
subthreshold ANCs at fixed \(C_0\).

The effect of changing the subthreshold ANCs is much larger than the
effect of moderate variations of \(C_0\).  The contribution of the
subthreshold \(1^{-}\) state to the total \(E1\) \(S\) factor at
\(300\) keV is about \(71\)--\(74\%\).  The remaining contribution comes
mainly from interference between the subthreshold \(1^{-}\) state and
the broad \(1^{-}\) resonance at \(E_x=9.585\) MeV.  The direct \(E1\)
capture contribution to the ground state is negligible even for
\(C_0=740~{\rm fm}^{-1/2}\).  For the \(E2\) transition, the constructive
interference required by the fit between the subthreshold resonance and
direct-capture amplitudes increases \(S_{E2}(300~{\rm keV})\) compared
with the pure subthreshold contribution
\(S_{E2}^{({\rm SR})}(300~{\rm keV})\).  The renormalized subthreshold
ANCs of Ref.~\cite{Hebborn} reduce both
\(S_{E1}^{(SR)}(300~{\rm keV})\) and \(S_{E2}^{(SR)}(300~{\rm keV})\) by about
\(30\%\) compared with the values obtained using the original
subthreshold ANCs of Refs.~\cite{Brune,Avila}.  Consequently, the ANC
update decreases the total low-energy \(S\) factor.

Having established the impact of the updated subthreshold ANCs and the
large ground-state ANC on the \(E1\) and \(E2\) transitions, we now
consider the total \(S\) factor,
\[
S_{\rm tot}(E)=S_{E1}(E)+S_{E2}(E)+S_{\rm casc}(E),
\]
where \(S_{\rm casc}(E)\) is the cascade contribution.  The
\(S_{E1}(E)\) and \(S_{E2}(E)\) contributions are taken from the fits to
the data of Ref.~\cite{Assunc}, while the cascade contribution is taken
from Ref.~\cite{deBoer}.

\begin{figure}[t]
\centering
\includegraphics[width=0.7\linewidth,angle=-90]{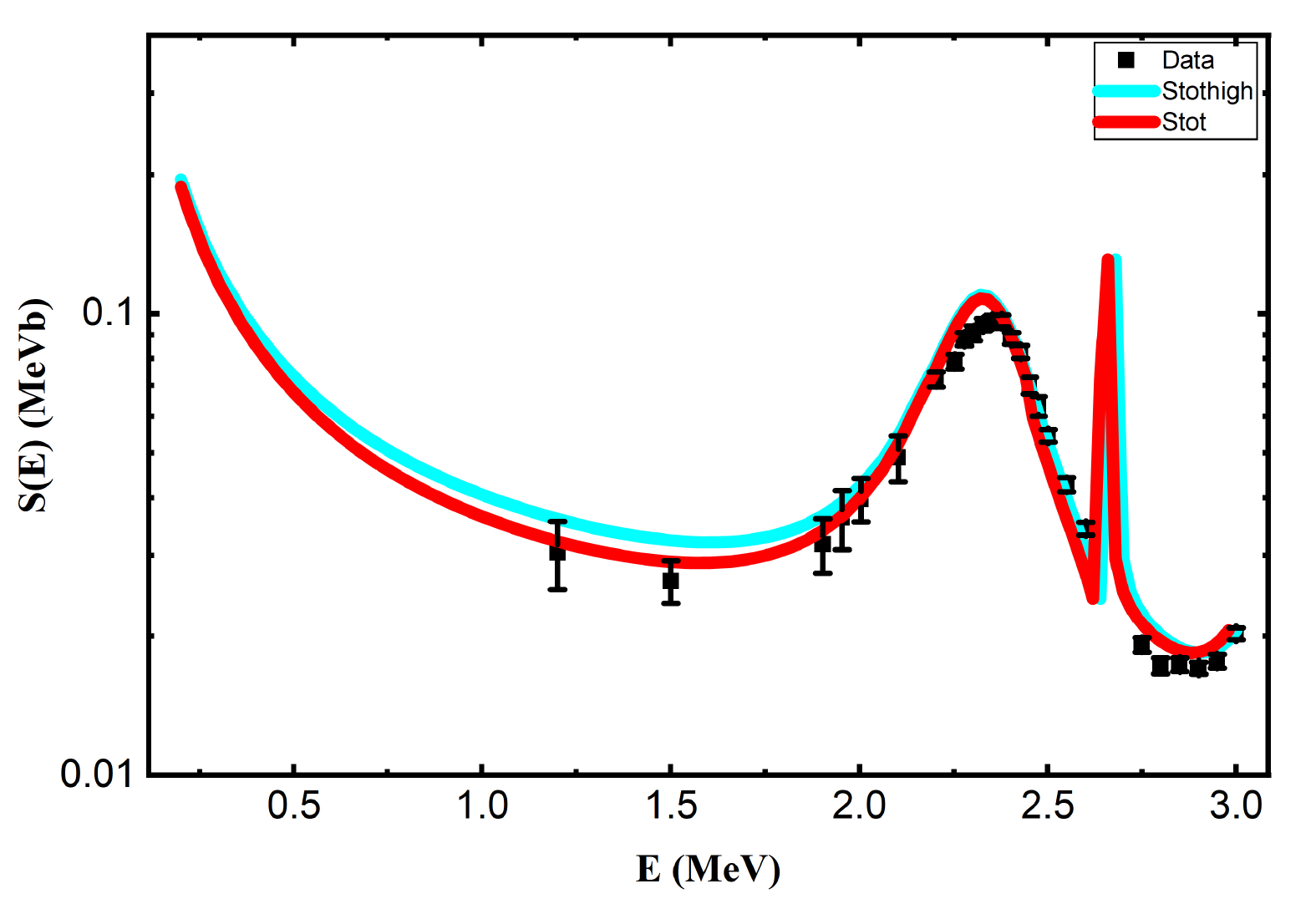}
\caption{
The total \(S_{\rm tot}(E)\) factor obtained from the present \(E1\) and
\(E2\) fits.  The red solid line is obtained by summing the red solid
curves in Figs.~\ref{fig_SE1} and \ref{fig_SE2}, together with the
cascade contribution.  It corresponds to
\(C_1=1.83\times10^{14}~{\rm fm}^{-1/2}\),
\(C_2=0.98\times10^5~{\rm fm}^{-1/2}\), and
\(C_0=740~{\rm fm}^{-1/2}\).  For comparison, the green solid line shows
the result obtained with \(C_0=900~{\rm fm}^{-1/2}\).
}
\label{fig_Stot}
\end{figure}

The calculated \(S_{\rm tot}(E)\) is compared with the total-capture
data of Refs.~\cite{Schurmann,Yamaguchi}.  The data from Ref.~\cite{Schurmann}   
  are supplemented by the two low-energy total \(S\)-factor points at
\(E=1.2\) and \(1.5\) MeV from Ref.~\cite{Yamaguchi}.  The result is
shown in Fig.~\ref{fig_Stot}.

The ANCs of the subthreshold \(1^{-}\) and \(2^{+}\) states, the
ground-state ANC, and the available low-energy total data, especially
the points of Ref.~\cite{Yamaguchi}, impose strong restrictions on the
physically acceptable range of \(S(300~{\rm keV})\).

It is important to distinguish between the quality of the separate
multipole fits and the quality of the comparison with the total-capture
data.  The \(E1\) and \(E2\) contributions were adjusted and tested
against the corresponding extracted multipole data of Ref.~\cite{Assunc}.
In that comparison, the individual calculated multipole curves provide a
very good description of the data.  The total \(S\)-factor comparison,
however, involves a different experimental situation.  It combines the
calculated \(E1\), \(E2\), and cascade contributions and compares their
sum with total-capture data from Ref.~\cite{Schurmann} and with the two
low-energy total points of Ref.~\cite{Yamaguchi}.

Therefore, one cannot expect a perfect fit to the total data of
Refs.~\cite{Schurmann,Yamaguchi}.  These measurements may have different
absolute normalizations, quoted uncertainties, systematic effects, and
correlations from those entering the extracted multipole data of
Ref.~\cite{Assunc}.  Consequently, the total comparison should be viewed
as a consistency test of the complete \(E1+E2+\)cascade calculation
rather than as a direct refit of the individual multipole components.
Nevertheless, the calculated total \(S\) factor obtained from the
multipole fits to the data of Ref.~\cite{Assunc} shows good overall
agreement with the independent total-capture data of
Refs.~\cite{Schurmann,Yamaguchi}.

For the adopted ANC values
\(C_1=1.83\times10^{14}~{\rm fm}^{-1/2}\),
\(C_2=0.98\times10^{5}~{\rm fm}^{-1/2}\), and the posterior median value
\(C_0^{\rm med}=740~{\rm fm}^{-1/2}\), we obtain
\[
S_{\rm tot}(300~{\rm keV})=118~{\rm keV\,b}.\]
Propagating the \(S_{E2}\)-only posterior for \(C_0\), with
\[
688\le C_0\le 798~{\rm fm}^{-1/2}
\qquad (68\%~{\rm credible~interval}),
\]
gives the corresponding range
\begin{align}
S_{\rm tot}(300~{\rm keV})
\simeq 116\text{--}119~{\rm keV\,b}
\label{C0uncinterval}
\end{align}
Thus, the Bayesian inference strongly constrains the uncertainty in
\(S_{\rm tot}(300~{\rm keV})\) associated with the uncertainty in
\(C_0\).

In addition to the uncertainty caused by \(C_0\), one must also account
for the more sensitive uncertainties associated with the subthreshold
ANCs \(C_1\) and \(C_2\).  According to Ref.~\cite{Hebborn},
$C_1=(1.83\text{--}1.84)\times10^{14}~{\rm fm}^{-1/2}$  and
$C_2=(0.98\text{--}1.07)\times10^5~{\rm fm}^{-1/2}.$
These ranges give
\begin{align}
S_{E1}(300\,{\rm keV})=62\text{--}63~{\rm keV\,b},
\label{SE1}
\end{align}
and
\begin{align}
S_{E2}(300\,{\rm keV})=50\text{--}54~{\rm keV\,b}.
\label{SE2}
\end{align}
Thus, before adding the cascade contribution,
\[
S_{E1}(300\,{\rm keV})+S_{E2}(300\,{\rm keV})
=
112\text{--}117~{\rm keV\,b}.
\]
After including the fixed cascade contribution, this corresponds to
\[
S_{\rm tot}(300\,{\rm keV})=117\text{--}122~{\rm keV\,b}.
\]

Taking the envelope of the ANC-driven uncertainties from both the
\(C_0\) posterior and the adopted \(C_1,\,C_2\) ranges, we obtain
\[
S_{\rm tot}(300~{\rm keV})
\simeq
116\text{--}122~{\rm keV\,b}.
\]
Equivalently, for the central value
\[
S_{\rm tot}(300~{\rm keV})\simeq 118~{\rm keV\,b},
\]
this may be written as
\[
S_{\rm tot}(300~{\rm keV})
=
118^{+4}_{-2}~{\rm keV\,b}.
\]
Here the quoted uncertainty reflects the propagated ANCs uncertainty
only; it does not include additional possible systematic uncertainties
from the choice of \(R\)-matrix parametrization, channel radius,
higher-lying levels, or experimental normalization effects.

If all these additional uncertainties are included, the total error band
should be larger than the ANC-driven range quoted above.  A conservative
estimate can be obtained by adopting a relative uncertainty comparable to
that of the deBoer {\it et al.} \cite{deBoer}, and applying the same
relative uncertainty to our central value
\(S_{\rm tot}(300~{\rm keV})=118~{\rm keV\,b}\), gives
\[
\Delta S_{\rm tot}
\simeq
118\,\frac{20}{140}
\simeq
17~{\rm keV\,b}.
\]
Thus, as a conservative systematic estimate, one may write
\begin{align}
S_{\rm tot}(300~{\rm keV})
\simeq
118\pm 17~{\rm keV\,b}.
\label{Stotuncerint}
\end{align}
This broader uncertainty should be interpreted as an approximate
systematic scale rather than as the Bayesian uncertainty obtained from
the \(C_0\) posterior alone.

The main conclusion is that the latest updates of the subthreshold and
ground-state ANCs push \(S(300~{\rm keV})\) noticeably downward compared
with the value obtained in Ref.~\cite{deBoer}.

\section{Physical relation between \(S(300~{\rm keV})\) and the lower edge of the black-hole mass gap}
\label{sec:S300_BHgap}

From the standpoint of stellar-evolution physics, the relation between
the lower edge of the first-generation BH mass gap and
\(S(300~{\rm keV})\) should be inverse.  Increasing the
\(^{12}{\rm C}(\alpha,\gamma)^{16}{\rm O}\) rate enhances the conversion
of \(^{12}{\rm C}\) into \(^{16}{\rm O}\) during core-helium burning and
therefore leaves a smaller residual carbon abundance at helium
exhaustion.  A smaller post-helium-burning carbon abundance weakens
subsequent shell-carbon burning, reduces its stabilizing effect on the
core, and promotes an earlier onset of pulsational pair-instability episodes 
or a full pair-instability supernova behavior.  Consequently, increasing
\(S(300~{\rm keV})\) should lower, rather than raise, the maximum mass
of a first-generation BH.  Equivalently, larger values of
\(S(300~{\rm keV})\) should correspond to a smaller lower edge of the
black-hole mass gap.

This inverse trend is physically natural and must remain the basis of
any consistent interpretation of gravitational-wave data.  If very large
values of \(S(300~{\rm keV})\) shift the lower edge of the
pair-instability mass gap down to \(\sim40\)--\(50\,M_\odot\), then BHs
observed in the \(50\)--\(70\,M_\odot\) range would lie inside the
predicted gap rather than below it.
Therefore, if very large values of \(S(300~{\rm keV})\) shift the lower
edge of the pair-instability mass gap down to \(\sim 40\)--\(50\,M_\odot\),
then BHs observed in the \(50\)--\(70\,M_\odot\) range would lie
inside the predicted gap rather than below it.  

This difficulty is not only astrophysical.  Very large inferred
values of \(S(300~{\rm keV})\), required to push the lower edge of the
mass gap down to \(\sim40\)--\(50\,M_\odot\), are also incompatible with
the experimental nuclear data for
\(^{12}{\rm C}(\alpha,\gamma)^{16}{\rm O}\) and with contemporary ANC
measurements.  Agreement with a population-level gravitational-wave
inference alone is therefore insufficient; the inferred reaction rate
must also satisfy the independent nuclear-physics constraints.

It is important to emphasize that the lower edge of the black-hole mass
gap is not directly measured.  It is inferred from the observed
black-hole population, and the result depends on the assumptions made in
the analysis.  These assumptions include how many BHs may come
from hierarchical mergers, what spin and mass distributions are used,
how observational selection effects are treated, and what population
model is adopted.

For this reason, different studies do not always determine the same
quantity.  Some analyses find evidence for BHs in the
\(50\)--\(70\,M_\odot\) range and therefore favor a higher lower edge of
the gap.  Other analyses allow a lower edge near
\(40\)--\(50\,M_\odot\), but this conclusion depends strongly on the
chosen population model and on how second-generation BHs are
treated.  Thus these results should not be regarded as direct
measurements of one unique, well-defined lower edge of the
first-generation pair-instability mass gap.

Thus, purely astrophysical population inferences from current
gravitational-wave data do not robustly determine the lower edge of the
black-hole mass gap in a model-independent way.  Independent
nuclear-physics information must be included from the outset.

The essential nuclear inputs for determining the maximum mass of
first-generation BHs, and hence the lower edge of the
pair-instability mass gap, include the triple-\(\alpha\) reaction rate,
the \(^{12}{\rm C}+{}^{12}{\rm C}\) fusion rate, and the
\(^{12}{\rm C}(\alpha,\gamma)^{16}{\rm O}\) reaction rate at
astrophysically relevant temperatures.  The triple-alpha reaction first produces the $^{12}{\rm C}$ nuclei during  helium burning. These carbon nuclei then serve as the target for the
subsequent $^{12}{\rm C}(\alpha,\gamma)^{16}{\rm O}$ reaction, which
converts part of the newly formed carbon into oxygen.  The reaction rate for the triple-alpha process was adopted from Ref.~\cite{Caughlan}.

The role of later carbon and oxygen burning should also be kept in
perspective.  Work~\cite{Farmer20} considered variations of
the triple-alpha reaction, \(^{12}{\rm C}(\alpha,\gamma)^{16}{\rm O}\),
\(^{12}{\rm C}+{}^{12}{\rm C}\), and
\(^{16}{\rm O}+{}^{16}{\rm O}\), but found that the dominant nuclear
sensitivity of the black-hole mass-gap boundary comes from
\(^{12}{\rm C}(\alpha,\gamma)^{16}{\rm O}\).  The triple-alpha reaction
is also important because it produces \(^{12}{\rm C}\) during helium
burning, while \(^{12}{\rm C}(\alpha,\gamma)^{16}{\rm O}\) converts
part of this carbon into \(^{16}{\rm O}\).  Thus the final carbon
abundance is governed mainly by the competition between these two
helium-burning reactions.

This conclusion was confirmed in \cite{Mehta22} using updated
stellar-evolution calculations: the lower edge of the mass gap changes
strongly with the \(^{12}{\rm C}(\alpha,\gamma)^{16}{\rm O}\) rate,
whereas the later heavy-ion fusion rates
\(^{12}{\rm C}+{}^{12}{\rm C}\),
\(^{12}{\rm C}+{}^{16}{\rm O}\), and
\(^{16}{\rm O}+{}^{16}{\rm O}\) move the boundary by only
\(\lesssim 1\,M_\odot\).  Therefore, for fixed stellar-evolution
inputs, such as metallicity, mass loss, mixing, and the adopted
triple-\(\alpha\) and carbon-burning rates
\cite{Farmer20,Woosley21,Mehta22}, the leading nuclear uncertainty is
the \(^{12}{\rm C}(\alpha,\gamma)^{16}{\rm O}\) rate, because it fixes
the residual carbon abundance after core-helium burning.

Therefore, any inference that maps the observed black-hole mass
distribution onto a value of \(S(300~{\rm keV})\) must remain consistent
with independent nuclear constraints on the
\(^{12}{\rm C}(\alpha,\gamma)^{16}{\rm O}\) reaction, including direct
and indirect measurements, \(R\)-matrix extrapolations, and ANC
constraints on the subthreshold and ground-state contributions.

In Ref.~\cite{Mehta22}, \(\sigma=0\) corresponds to the adopted central
value
\[
S_{0}(300~{\rm keV})=140~{\rm keV\,b},
\]
and one unit change in \(\sigma\) corresponds approximately to a
variation of \(20~{\rm keV\,b}\).  Thus, the shifted value may be written
as
\[
S(300~{\rm keV})\simeq S_0(300~{\rm keV})+20\sigma~{\rm keV\,b}.
\]
For example, the \(\sigma=-1\) point corresponds approximately to
\[
S(300~{\rm keV})\simeq120~{\rm keV\,b}.
\]
The present ANC-constrained value,
\[
S_{\rm tot}(300~{\rm keV})=118~{\rm keV\,b},
\]
therefore corresponds to
\[
\sigma\simeq -1.1 .
\]
Thus, the present ANC-constrained rate is very close to the
\(-1\sigma\) rate inferred from Refs.~\cite{deBoer,Mehta22}, differing
from it by only about \(2\%\).

\begin{figure}[t]
\centering
\includegraphics[width=1.0\linewidth]{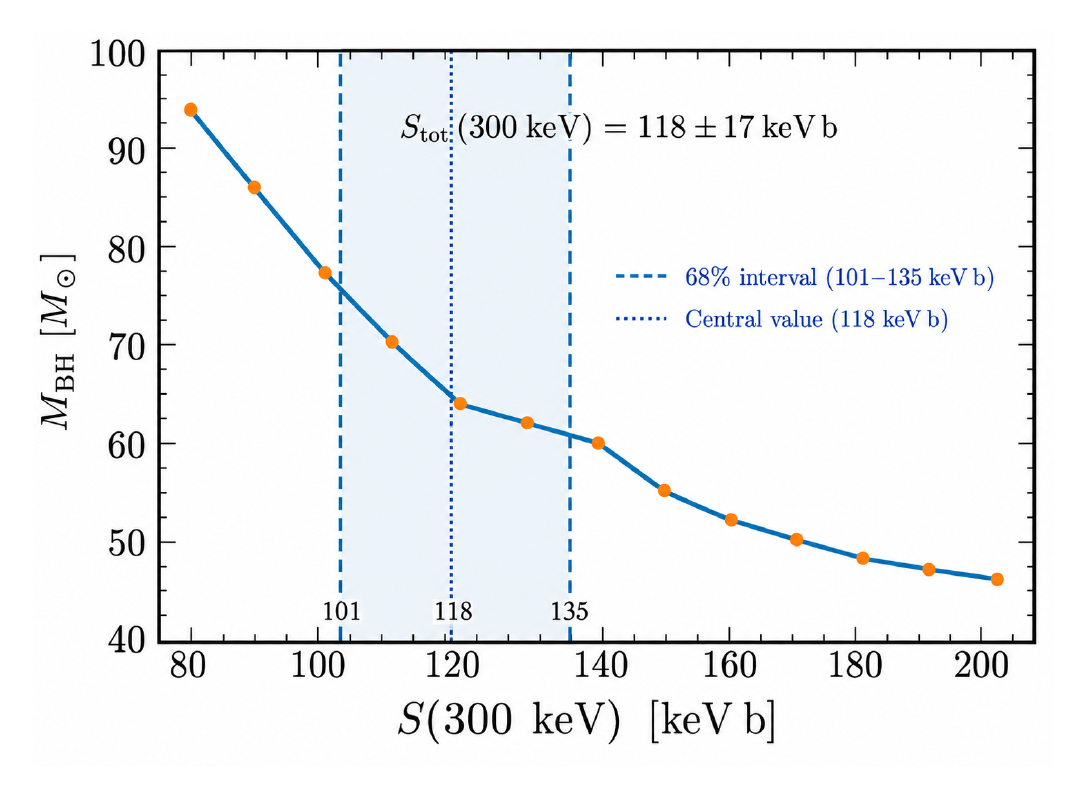}
\caption{
Dependence of the maximum first-generation black-hole mass, identified
with the lower edge of the pair-instability black-hole mass gap, on the
astrophysical \(S\) factor \(S(300~{\rm keV})\) for the
\(^{12}{\rm C}(\alpha,\gamma)^{16}{\rm O}\) reaction.  The curve was
redrawn from the lower-edge result of Ref.~\cite{Mehta22}.  The original
horizontal axis in Ref.~\cite{Mehta22} was expressed in terms of the
reaction-rate variation parameter
\(\sigma[^{12}{\rm C}(\alpha,\gamma)^{16}{\rm O}]\).  Here it is
converted approximately to \(S(300~{\rm keV})\), where
\(\sigma=0\) corresponds to \(S_{0}(300~{\rm keV})=140~{\rm keV\,b}\), and
one unit change in \(\sigma\) corresponds approximately to
\(20~{\rm keV\,b}\).  The figure illustrates the inverse trend: larger
values of \(S(300~{\rm keV})\) correspond to a smaller maximum
first-generation black-hole mass and hence to a lower edge of the mass
gap.
}
\label{fig_MBH}
\end{figure}

To make this connection more transparent, we transformed the dependence
shown in Ref.~\cite{Mehta22} from the reaction-rate variation parameter
\(\sigma\) to an approximate dependence on
\(S(300~{\rm keV})\).  
This transformation allows the lower-edge black-hole mass
\(M_{\rm BH}/M_\odot\) to be displayed directly as a function of
\(S(300~{\rm keV})\), as shown in Fig.~\ref{fig_MBH}.

The resulting curve should be understood as an approximate
reparametrization of the horizontal axis of Ref.~\cite{Mehta22}, not as
a new stellar-evolution calculation.  Using this curve together with the
\(S(300~{\rm keV})\) interval obtained in Eq.~(\ref{Stotuncerint}), we
infer the corresponding interval for the maximum first-generation
black-hole mass, or equivalently the lower edge of the pair-instability
mass gap,
\begin{align}
\frac{M_{\rm BH}}{M_\odot}
\simeq 61\text{--}75 .
\label{final_BHmass}
\end{align}

\begin{figure}[t]
\centering
\includegraphics[width=0.85\linewidth]{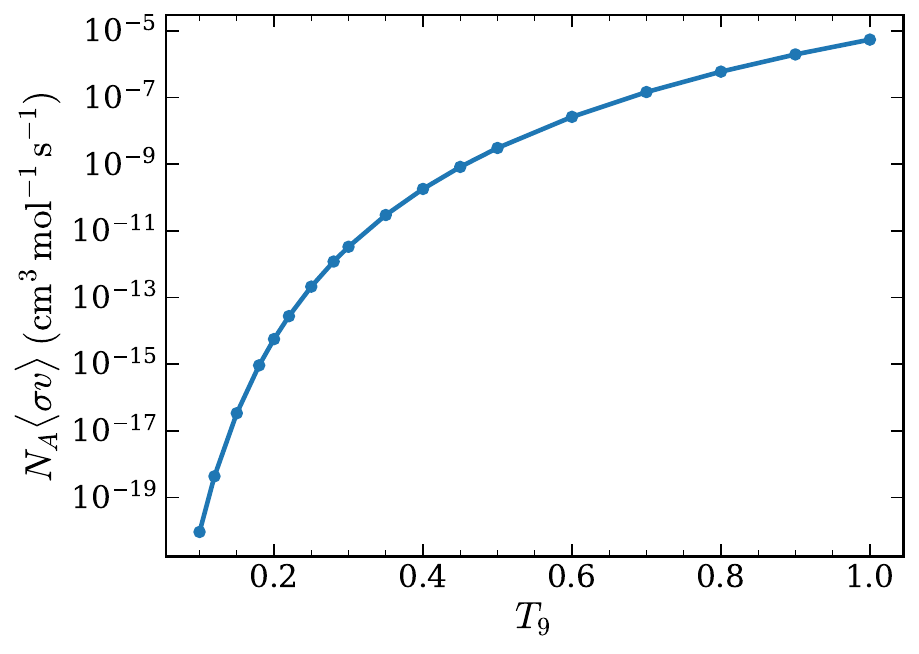}
\caption{
Present ANC-constrained thermonuclear reaction rate
\(N_A\langle\sigma v\rangle\) for the
\(^{12}{\rm C}(\alpha,\gamma)^{16}{\rm O}\) reaction as a function of
temperature \(T_9\), where \(T_9\) is the temperature in units of
\(10^9\) K.
}
\label{fig_RecRate}
\end{figure}

The present ANC-constrained reaction rate was calculated from the
total \(S_{\rm tot}(E)\) obtained in this work using the integration
interval \(0.01\le E\le2.0~{\rm MeV}\).  The resulting
thermonuclear reaction rate \(N_A\langle\sigma v\rangle\) is shown in
Fig.~\ref{fig_RecRate}.

\begin{figure}[t]
\centering
\includegraphics[width=0.85\linewidth]{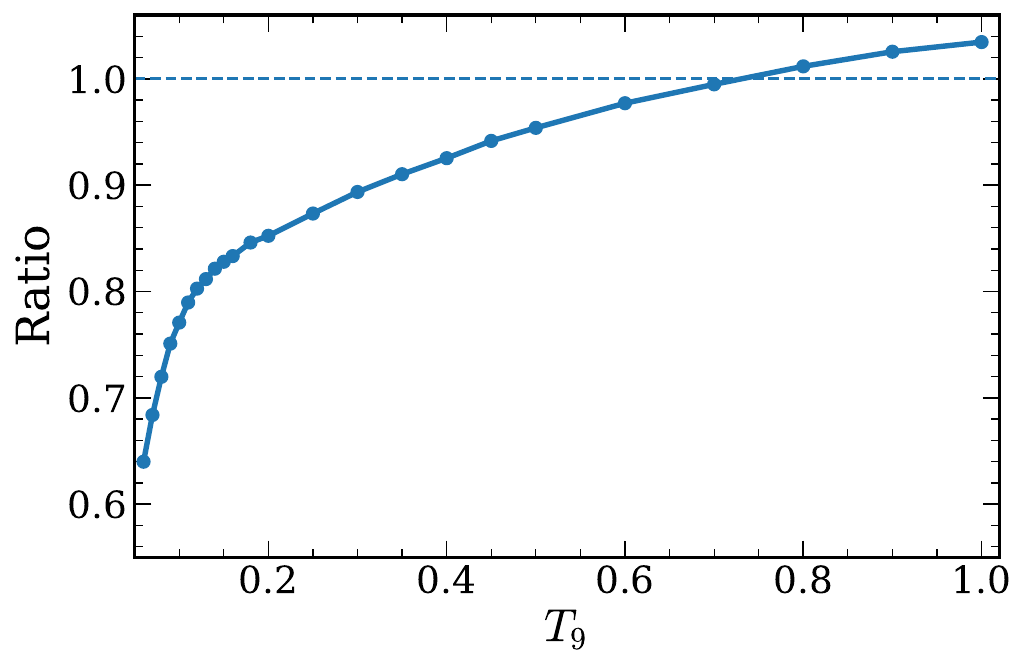}
\caption{
Ratio of the present ANC-constrained thermonuclear reaction rate to the
adopted rate from Table XXV of Ref.~\cite{deBoer}.
}
\label{fig_RecRateRatio}
\end{figure}

The ratio of the present rate to the adopted rate from
Table XXV of Ref.~\cite{deBoer} is shown in
Fig.~\ref{fig_RecRateRatio}.  At low temperatures, where the Gamow
window lies in the sub-Coulomb region, the reaction rate is controlled
mainly by the tails of the subthreshold states and therefore by the
corresponding ANCs.  In this region the ratio is close to the reduction
expected from the \(S\)-factor ratio
\begin{align}
\frac{S_{\rm present}(300~{\rm keV})}
     {S_{\rm deBoer}(300~{\rm keV})} =
\frac{118}{140}
\simeq0.843 .
\label{Ratio}
\end{align}
Indeed, near the core-helium-burning temperature \(T_9\simeq0.2\), the
direct rate comparison gives
\[
\frac{R_{\rm present}}{R_{\rm deBoer}}\simeq0.85 .
\]
As the temperature increases, the Gamow window moves to higher energies,
where the subthreshold tails are less dominant and the measured
\(S(E)\) data constrain the rate more strongly.  Since both the present
calculation and the deBoer evaluation describe the experimental data in
this higher-energy region, the ratio gradually approaches unity.

\section{Conclusion}

The central result of this work is that the extrapolated
\(^{12}{\rm C}(\alpha,\gamma)^{16}{\rm O}\) \(S\) factor at
\(E=300\) keV must remain constrained by nuclear physics and cannot be
determined solely from BH population arguments.  The quantity
\(S(300~{\rm keV})\) is directly connected with the ANCs of the
subthreshold and ground-state configurations, in particular
\(C_1\), \(C_2\), and \(C_0\).  By contrast, the lower edge of the
black-hole mass gap is not directly measured; it is inferred
indirectly from gravitational-wave population analyses and depends on
assumptions about spin distributions, hierarchical mergers, selection
effects, priors, and the adopted population model.  Therefore, an
astrophysical inference that requires values of \(S(300~{\rm keV})\)
outside the independently allowed nuclear-physics domain is not
physically well constrained and is difficult to reconcile with
independent nuclear constraints.

The available ANC information does not support very large values such as
\[
S(300~{\rm keV})\sim 240\text{--}263~{\rm keV\,b},
\]
which appear in some interpretations based on a very low adopted
mass-gap edge.  Even values near
\(S(300~{\rm keV})\simeq170~{\rm keV\,b}\) are difficult to reconcile
with present ground-state ANC constraints.  In this work we therefore
impose the independently determined ANC constraints from the outset.
The subthreshold ANCs \(C_1\) and \(C_2\) are taken from sub-Coulomb
transfer analyses, where the reaction is peripheral and the cross
section is governed primarily by the relevant ANCs and the
\(^{6}{\rm Li}\) ANC.  The latter is independently constrained by
six-body ab initio calculations and by the excellent agreement with the
LUNA measurement of
\(\alpha+d\rightarrow{}^{6}{\rm Li}+\gamma\) \cite{Anders14}.  For the ground-state ANC
we use the modern range summarized in Table~\ref{Table_C0}, which favors
\(C_0\gtrsim600~{\rm fm}^{-1/2}\) and is incompatible with the very small
ground-state ANC used in the older evaluation.

With these nuclear constraints imposed, the calculated total \(S\)
factor is not an unconstrained fit to the measured total data.  Rather,
it is an ANC-constrained \(R\)-matrix extrapolation.  The
\(S_{E2}\)-only Bayesian analysis gives a posterior maximum near
\[
C_0^{\rm MAP}\simeq703~{\rm fm}^{-1/2},
\]
with a 68\% credible interval
\[
688\le C_0\le798~{\rm fm}^{-1/2}.
\]
This posterior does not support a drift toward small values of
\(C_0\).  The resulting total extrapolated value is
\[
S_{\rm tot}(300~{\rm keV})\simeq 118~{\rm keV\,b}.
\]
Propagating the ANC-driven uncertainty gives
\[
S_{\rm tot}(300~{\rm keV})
\simeq 116\text{--}122~{\rm keV\,b},
\]
or equivalently
\[
S_{\rm tot}(300~{\rm keV})
=
118^{+4}_{-2}~{\rm keV\,b},
\]
where the quoted uncertainty reflects the propagated ANC uncertainty
only.  Additional systematic uncertainties associated with the
\(R\)-matrix parametrization, channel radius, higher-lying levels, and
experimental normalizations should be treated separately.

The measured total \(S\)-factor data alone do not uniquely determine
\(S_{\rm tot}(300~{\rm keV})\). Different ANC choices can reproduce the measured energy region while leading to different low-energy extrapolations.
 Therefore, a Bayesian analysis based only on the total
\(S\)-factor data is premature as a determination of
\(S_{\rm tot}(300~{\rm keV})\). A physically meaningful Bayesian treatment should therefore propagate independently established ANC uncertainties rather than freely
determining the ANCs from the present total-capture data alone.

Finally, using the transformed relation between the lower edge of the
black-hole mass gap and \(S(300~{\rm keV})\), based on
Ref.~\cite{Mehta22}, the present ANC-constrained range gives
\[
\frac{M_{\rm BH}}{M_\odot}\simeq61\text{--}75 .
\]
This estimate is consistent with the mass-gap boundary inferred in
Refs.~\cite{Farmer20,Woosley21,Mehta22}, where a representative value
\(59^{+34}_{-13}\,M_\odot\) was obtained, and lies within that interval. It also overlaps with the interval \(50\text{--}70\,M_\odot\)  \cite{Wang25}.

Thus, the nuclear-physics constrained value of \(S(300~{\rm keV})\)
favors a relatively high lower edge of the first-generation black-hole
mass gap.  In particular, the present result does not support a strong
downward shift of this boundary to \(\sim45\,M_\odot\), which would
require very large \(S(300~{\rm keV})\) values.  Agreement with gravitational-wave population observables alone is therefore insufficient; the inferred reaction rate must also remain consistent with  independent nuclear-physics constraints.

\end{document}